\documentclass{aa}
\usepackage[final]{graphicx}
\usepackage{txfonts}

\begin{document}

\title{Investigating the EGRET-radio galaxies link with $INTEGRAL$:\\
the case of $3$EG~J$1621+8203$ and NGC~$6251$\thanks{Based on  observations with INTEGRAL,
an ESA mission with instruments and science data center funded  by ESA member states (especially the PI
countries: Denmark, France, Germany, Italy, Switzerland, Spain),  Czech Republic and Poland, and with the
participation of Russia and the USA.}}

\author{L.~Foschini\inst{1}, M. Chiaberge\inst{2},  P. Grandi\inst{1}, I.A. Grenier\inst{3,4}, M. Guainazzi\inst{5}, 
W. Hermsen\inst{6,7}, G.G.C. Palumbo\inst{8}, J. Rodriguez\inst{4,9}, S. Chaty\inst{3,4}, S. Corbel\inst{3,4},
G.~Di~Cocco\inst{1}, L. Kuiper\inst{6}, G.~Malaguti\inst{1}}

\offprints{L. Foschini  - \email{foschini@bo.iasf.cnr.it}}

\institute{Istituto di Astrofisica Spaziale e Fisica Cosmica (IASF) del CNR/INAF - Sezione di Bologna,
Via Gobetti 101, 40129 Bologna (Italy)
\and
Istituto di Radioastronomia (IRA) del CNR/INAF, Via Gobetti 101, 40129 Bologna (Italy)
\and
Universit\'e Paris VII Denis-Diderot 2 place Jussieu 75251 Paris Cedex 05 (France)
\and
CEA Saclay, DSM/DAPNIA/SAp (CNRS FRE 2591) F$-91191$ Gif-sur-Yvette Cedex, (France)
\and
XMM-Newton SOC - European Space Astronomy Center of ESA, Vilspa, Apartado 50727, E-28080 Madrid (Spain)
\and
SRON National Institute for Space Research, Sorbonnelaan 2, 3584 CA Utrecht (The Netherlands)
\and
Astronomical Institute ``Anton Pannekoek'', University of Amsterdam, Kruislaan 403, 1098 SJ Amsterdam, (The Netherlands).
\and
Dipartimento di Astronomia, Universit\`a di Bologna, Via Ranzani 1, 40127 Bologna (Italy)
\and
INTEGRAL Science Data Centre, Chemin d'\'Ecogia 16, Versoix (Switzerland)
}

\date{Received 12 November 2004; accepted 6 December 2004}

\abstract{The analysis of an \emph{INTEGRAL} AO2 observation of the error contours of the EGRET source  3EG~J$1621+8203$ is presented. The only
source found inside the error contours for energies between 20 and 30 keV at $5\sigma$  detection significance is the FR I radio galaxy
NGC~$6251$. This supports the identification of NGC~$6251$ with  3EG~J$1621+8203$. The observed flux  is higher and softer than observed
in the past, but consistent with a variable blazar-like spectral energy distribution.

 }

\authorrunning{L. Foschini et al.}
\titlerunning{Investigating the EGRET-radio galaxies link with INTEGRAL}

\maketitle

\section{Introduction}
One of the long standing problems of modern high--energy astrophysics is to understand the
nature of sources emitting the highest energy $\gamma-$rays.
The Energetic Gamma--Ray Experiment Telescope (EGRET) Catalog, obtained from high--energy $\gamma-$ray observations ($E > 100$ MeV)
performed between $1991$ and $1995$ (Hartman et al. 1999),  changed our understanding of the $\gamma$-ray sky. The third version of this
catalog contains $271$ point sources and about 150 of them are still unidentified. The search for the counterparts of these
sources is particularly challenging since the point spread function (PSF) of EGRET is about $6^{\circ}$ at $100$ MeV (FWHM) and the error
contours are large, typically extent $0.5^{\circ}-1^{\circ}$. During recent years, much effort has been spent in gathering all the useful
data from the online multiwavelength archives, searching for possible counterparts. Despite some interesting results, 
the quest is still open (for a review see, e.g., Caraveo 2002, Grenier 2004, Mukherjee \& Halpern 2004, Thompson 2004).

The EGRET sources of extragalactic origin are mainly blazars. The correlation of EGRET  error boxes with flat--spectrum radio sources has
been revisited using deeper radio  interferometric surveys at $8.4$ GHz and improved radio spectral measurements in the northern and
southern sky, down to $\delta > -40^{\circ}$ (Sowards-Emmerd et al. 2003, 2004). These interferometric data have allowed us to discriminate
more efficiently between sources with a flat spectrum core from those with softer lobe emission. Combined with the associations proposed by Mattox et al.
(2001) using older surveys at $\delta < -40^{\circ}$, a total of 128 AGN counterparts have been proposed. $60$\% of the identifications
are considered very likely. Five associations with nearby Fanaroff-Riley type I and II radiogalaxies (FRI and FRII, respectively), namely
3EG~J$0459+3352$ (Sowards--Emmerd et al. 2003), 3EG~J$1324-4313$ (Cen A, Steinle et al. 1998), 3EG~J$1646-0704$ (Sowards--Emmerd et
al. 2004), 3EG~J$1736-2908$ (Di Cocco et al. 2004),  3EG~J$1735-1500$ (Combi et al. 2003) and 3EG~J$1621+8203$ (Mukherjee et al. 2002)
open the exciting possibility of studying misaligned blazars in gamma rays. While blazars can be powerful $\gamma$-ray emitters
because of relativistic beaming and amplification, the identification of $\gamma$-ray radiogalaxies poses interesting problems.

\begin{figure}[t]
\centering
\vspace{12pt}
\includegraphics[scale=0.45]{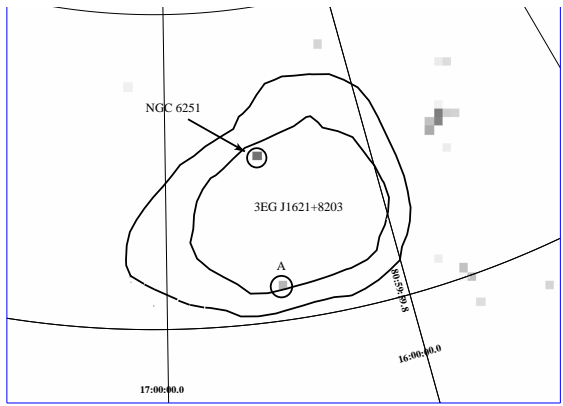}
\caption{IBIS/ISGRI deconvolved significance map of the 3EG~J$1621+8203$ sky
region (exposure of $424$~ks), in the $20-30$~keV energy band, with a S/N threshold set to
$3\sigma$. The 3EG~J$1621+8203$ probability contours at 95\% and 99\% confidence level are shown.
``A'' indicates the unknown excess at $\approx 3.5\sigma$. See the text for more details.}
\label{images}
\end{figure}

According to the unified scheme of AGN by Urry \& Padovani (1995), the transition from blazars to radio galaxies is based on a
combination of orientation and relativistic beaming. Specifically, radio galaxies of type FRI and BL Lac objects are similar AGN seen
with jet directions at large angles to the observer's view or close to it. On the other hand, FR II radio galaxies and flat  spectrum
radio quasars (FSRQ), share the same orientation. The detection by  EGRET of copious $\gamma-$rays from blazar jets implies
that FR I and FR II radio galaxies  are also intrinsically powerful $\gamma-$ray emitters, but their observation at large aspect  angle
should yield a much fainter flux and softer spectrum. This is indeed the case for Cen A (Steinle et al. 1998), that has a photon spectral
index in the EGRET energy range ($E>100$ MeV) of $\Gamma=2.40\pm 0.28$, that appears to be slightly softer than the average value for
blazars ($\Gamma=2.15\pm 0.04$), although still consistent within the measurement errors.  Cen A has a large inclination angle ($\sim
70^{\circ}$), so that it is difficult to explain the detection of emission in the EGRET energy range, but some explanations 
have been proposed (see the discussions in Sreekumar et al. 1999, and Mukherjee et al. 2002).

Therefore, it is very interesting to search for other radio galaxies that may be counterparts of EGRET sources. Their
identification with EGRET sources would have a large impact yielding, for instance, valuable constraints on the
collimation of $\gamma-$ray emission from the jet and the energy of the jet particles. In addition, it would have a significant impact on
population studies since the density of radio galaxies is far higher than that of blazars, and a significant reassessment of the origin
of the diffuse extragalactic $\gamma-$ray background would be required.

Among the different cases available, we focus our attention on 3EG~J$1621+8203$, that Mukherjee et al.
(2002) have associated with the FR I radio galaxy NGC~$6251$  ($z=0.02488$). They analyzed data from X--ray
satellites (\emph{ROSAT}, \emph{ASCA})  and by cross--correlating with the NRAO VLA Sky Survey (NVSS), they
found that, among the several sources found in the EGRET error contours, NGC~$6251$ is the most promising
candidate to be a high--energy $\gamma-$ray emitter. However, it should be noted that inside the EGRET error
contour there are several other X--ray and radio sources, some of them unidentified. Therefore, NGC~$6251$ was
considered the best candidate among the \emph{known} sources.

In order to resolve this doubt, we requested for a long observation with the IBIS telescope (Ubertini et al. 2003)
on board the \emph{INTEGRAL} satellite (Winkler et al. 2003).  IBIS has a large field of view of
$19^{\circ}\times 19^{\circ}$ at half response, with a high angular resolution of $12'$ sampled in $5'$
pixels in the low energy ($0.015-1$ MeV) detector ISGRI (Lebrun et al. 2003) and in $10'$ pixels in the high
energy ($0.175-10$ MeV) layer PICsIT (Di Cocco et al. 2003). The point source location accuracy (PSLA) of
ISGRI can be down to $1'$ for a $30\sigma$ detection (Gros et al. 2003). These characteristics make IBIS a
valuable instrument to search for hard X-ray ($E>20$~keV) counterparts in the large error contours of the
EGRET sources.

Here we report the analysis of the region containing the 3EG~J$1621+8203$ error contours and the hard X-ray
emission from NGC~$6251$.

Throughout the paper we adopted $H_0=70$ km s$^{-1}$ Mpc$^{-1}$.

\section{$INTEGRAL$ data analysis}
\emph{INTEGRAL} observed 3EG~J$1621+8203$ from June 17, 2004, $18^{\rm h} 28^{\rm m}$ to June 23, 2004,
$08^{\rm h} 13^{\rm m}$ UTC with a $5\times 5$ dither pattern. The region of interest was  about $4.6^{\circ}$
off axis, because the observation of the EGRET source  (Prop ID 0220018, PI Foschini) was amalgamated
with the observation of the  galaxy cluster Abell 2256 (Prop ID 0220020, PI Fusco Femiano), which was the
on-axis target.

The \emph{INTEGRAL} data analysis described in the present work was done with the most recent
version of the Offline Standard Analysis (\texttt{OSA 4.1}\footnote{Available through the \emph{INTEGRAL Science
Data Centre} (Courvoisier et al. 2003) at \texttt{http://isdc.unige.ch/index.cgi?Soft+download}}) whose
algorithms for IBIS are described in Goldwurm et al. (2003). 

The spectrometer SPI (Vedrenne et al. 2003) was performing an annealing and therefore no data are available.
In addition, given the off-axis position, the source was not visible in the field of views of the X-ray monitor 
JEM-X (Lund et al. 2003) and the optical monitor OMC (Mas-Hesse et al. 2003), both also aboard \emph{INTEGRAL}.

The analysis was performed on all the available data for an effective exposure toward 3EG~J$1621+8203$ of
$424$ ks. The only known source  detected (signal--to--noise ratio $S/N=5\sigma$) inside the EGRET  error
contours is NGC~$6251$, found at coordinates (J2000) $\alpha=16^{\rm h} 33^{\rm m} 10^{\rm s}$ and
$\delta=+82^{\circ} 35' 46''$ with an uncertainty of $5'$ at the 90\% confidence level (Fig. \ref{images}).
The count rate in the $20-30$~keV energy band is $0.11\pm 0.02$ ($1\sigma$ error) corresponding to a
flux\footnote{The conversion of count rates  in physical units has been done by normalizing to the count
rates of the Crab Nebula,  obtained in several public calibration observations, and assuming  $F_{\rm
Crab}(E)=9.6\cdot E^{-2.1}$~ph~cm$^{-2}$~s$^{-1}$~keV$^{-1}$.}  of $(6.7\pm 2.1)\times
10^{-12}$~erg~cm$^{-2}$~s$^{-1}$ (90\% confidence level). There is no detection in higher energy bands, with an upper limit
($3\sigma$) of $0.05$ counts~s$^{-1}$ in the $30-40$~keV energy band, corresponding to a flux of  $4\times
10^{-12}$~erg~cm$^{-2}$~s$^{-1}$ assuming a Crab-like spectrum. 

Although the present statistics prevent a detailed spectral analysis, we performed some simulations with \texttt{xspec} (v. 11.3.1) and
the IBIS/ISGRI RMF/ARF matrices available in the \texttt{OSA 4.1} to understand the flux  differences between \emph{INTEGRAL} and
\emph{BeppoSAX} observations.  The flux and photon index values observed by \emph{BeppoSAX} corresponds to a count rate of about
$0.03$~c/s in the IBIS/ISGRI detector ($20-30$~keV energy range), to be compared with the value of $0.11\pm 0.02$ from
the present observation. This confirms that the flux variations are real.

Inside the EGRET probability contour there is only one other excess (marked with ``A'' in Fig.~\ref{images}) with
$S/N \approx 3.5\sigma$, but it does not correspond to any known source or even the unidentified  X-ray sources
found by Mukherjee et al. (2002). The closest source is 1RXS~J$163121.2+812641$, $8'$ distant from 
the excess A. The low $S/N$ is not sufficient to claim a significant source detection, but it is rather
likely to be a residual artifact of the deconvolution process.

The other detector layer of IBIS, PICsIT, had an overall exposure of $302$ ks, but no detection was found,
with an upper limit ($3\sigma$) of $1.2\times 10^{-9}$~erg cm$^{-2}$ s$^{-1}$, in the $672-1036$~keV band in
the 3EG~J$1621+8203$ region.

\begin{table}[!t]
\caption{Fluxes in the $20-30$ keV energy band obtained in different
epochs. In the case of \emph{ASCA} and \emph{XMM-Newton}, the value
is extrapolated from the best fit model (see the notes for details).
Columns: 
(1) Satellite name;
(2) Date of the observation [DD--MM--YYYY];
(3) Photon index;
(4) Flux [$10^{-12}$~erg cm$^{-2}$ s$^{-1}$].
The uncertainties in the parameters are at the 90\% confidence level.}
\centering
\begin{tabular}{lcccc}
\hline
Detector & Date & $\Gamma$ & $F$ \\
(1)      & (2)  & (3)      & (4) \\
\hline
\emph{ASCA}$^{\mathrm{a}}$       & $28-10-1994$ & $1.8\pm 0.2$             & $0.50\pm 0.06$  \\
\emph{BeppoSAX}$^{\mathrm{b}}$   & $19-07-2001$ & $1.79\pm 0.06$           & $1.73\pm 0.07$  \\
\emph{XMM-Newton}$^{\mathrm{c}}$ & $26-03-2002$ & $1.91_{-0.05}^{+0.08}$   & $1.19_{-0.06}^{+0.08}$ \\
\emph{INTEGRAL}$^{\mathrm{d}}$   & $17-06-2004$ & $2.1$$^{\mathrm{e}}$     & $6.7\pm 2.1$ \\
\hline
\end{tabular}
\label{tab:comparison}
\begin{list}{}{}
\item[$^{\mathrm{a}}$] Sambruna et al. (1999). Flux extrapolated from the fit in the $0.6-10$~keV energy band.
\item[$^{\mathrm{b}}$] Guainazzi et al. (2003), Chiaberge et al. (2003). Flux extracted from the joint fit LECS, 
MECS, PDS, in the $0.1-200$~keV energy band. 
\item[$^{\mathrm{c}}$] Gliozzi et al. (2004). Flux extrapolated from the fit in the $0.4-10$~keV energy band.
\item[$^{\mathrm{d}}$] Present work. 
\item[$^{\mathrm{e}}$] The conversion from rate to flux was done by assuming a Crab-like spectrum. 
\end{list}
\end{table}

\begin{figure}
\centering
\includegraphics[scale=0.4]{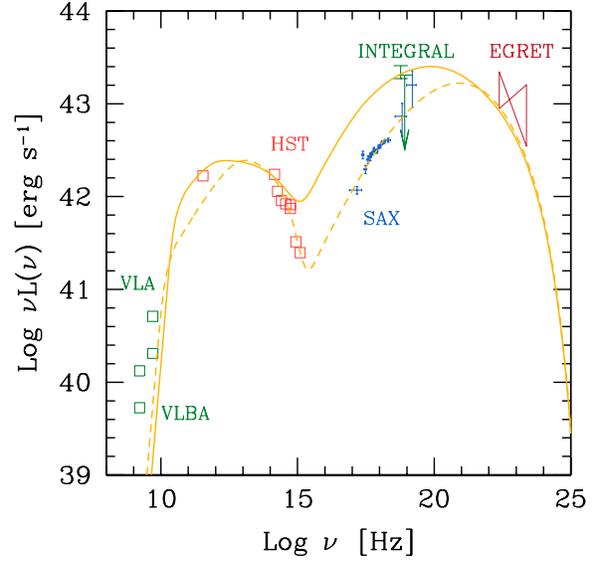}
\caption{Spectral energy distribution of NGC~$6251$ from Chiaberge et al. (2003) with the addition  of the \emph{INTEGRAL}/ISGRI data 
(horizontal bars are enlarged for clarity). The dashed line indicates the model fit to \emph{BeppoSAX} data; the continuous line represents 
the model fit to \emph{INTEGRAL} data. The new model parameters are as follows: source size  $R=3\times 10^{16}$~cm, injected electron luminosity $L_{\rm
inj}=6.7\times 10^{44}$~erg/s, slope of the injection $p=1.9$, $\gamma_{\rm min}=10^3$, $\gamma_{\rm max}=4\times 10^4$, magnetic field
$B=0.04$~G, escaping time $t_{\rm esc}=15R/c$, beaming factor $\delta=3.8$. See Chiaberge et al. (2003) for more details on the SSC model
parameters.}
\label{sed}
\end{figure}

\section{Discussion}
The identification of NGC~$6251$ as a counterpart of 3EG~J$1621+8203$ is a complex
puzzle, and with the present work we add some new pieces. To date Mukherjee et al. (2002) 
studied the \emph{ROSAT} ($0.1-2.4$ keV) and \emph{ASCA} ($0.6-10$ keV) observations,
and the radio survey NVSS at $20$ cm. They also investigated how the $\gamma-$ray
emission from a relativistic jet could decrease as a function of the observer's viewing angle
and found that by assuming an angle of $\theta=45^{\circ}$ for the jet axis of NGC~$6251$,
the flux emitted in the EGRET energy range is still detectable.

A complement, at radio wavelengths, to the work of Mukherjee et al. (2002) was recently published by
Sowards--Emmerd et al. (2003), with a survey at $8.4$ GHz, and they found that the only reliable counterpart
of 3EG~J$1621+8203$ is NGC~$6251$.

However, in addition to the spatial coincidence, it is necessary that the source is physically able to generate very high--energy
$\gamma-$rays. In this respect, another  piece in the puzzle was given by Guainazzi et al (2003) and Chiaberge et al. (2003): they found
that the spectral energy distribution (SED) of NGC~$6251$ has two peaks, very similar to the typical SED of blazars (Fossati et al.
1998). Chiaberge et al. (2003) found also an upper limit to the jet direction of $\theta < 18^{\circ}$ from the fit of the SED with the
synchrotron self--Compton (SSC) model (Fig.~\ref{sed}). On the other hand,   Jones \& Wehrle (2002) found $\theta<47^{\circ}$ with the
Very Long Baseline Interferometer (VLBI) measurement  of the jet--counterjet ratio. Moreover, Chiaberge et al. (2003) evaluated also the
beaming factor $\delta=3.2$,  that is lower than for blazars ($8\leq \delta \leq 23$), but greater than the case of Cen A
($\delta=1.2$). 

The SSC model establishes a link between the radio and the hard X-ray emission, but before the present work, the only observation at
energies greater than $10$ keV was performed by the BeppoSAX PDS. However, there could be some doubt about the flux evaluation with the
PDS,  since it is a collimated detector with a field of view (FOV) of $1.3^{\circ}$ and there is an active Seyfert nucleus at a distance
of $27'$ from NGC~$6251$ (see the source n. $17$ in the Fig. 3  of Mukherjee et al. 2002). Thanks to the unprecedented angular resolution
of IBIS/ISGRI,  it is possible to distinguish between the different contributions. The Seyfert nucleus  is not detected in the present
\emph{INTEGRAL} observation and we can exclude any contamination.  Moreover, IBIS/ISGRI shows that no other source or excess is present
in the error contours of EGRET down to a flux of  $\approx 5.4\times 10^{-12}$~erg~cm$^{-2}$~s$^{-1}$ ($4\sigma$, $20-30$~keV energy
range).

The SED of NGC~$6251$, updated with the data from the present observation, is shown in Fig.~\ref{sed}.   The model SED has been derived
using the numerical code described in Chiaberge \& Ghisellini (1999). The data at different wavelengths are
not obtained simultaneously, therefore, it is not possible to draw firm conclusions. However, with the \emph{INTEGRAL}
data it is possible to fit the data with the SSC, showing the source in a state with an overall higher flux. The parameters of the fit
are similar to those in Chiaberge et al.  (2003): in the present case the size of the source is slighlty smaller by about a factor of less
than 2 and the beaming factor increases from $\delta=3.2$ to $3.8$. Although the non-availability of simultaneous data prevents us from
performing a detailed spectral modeling, we speculate that the ``high state'' observed during the \emph{INTEGRAL} pointings might be produced
by a smaller region, located even closer to the jet base.

Another key question is still unexplained: the X-ray variability. It is indeed known that blazars
display strong variability on different  time scales, from hours to days (see e.g. Wagner \& Witzel 1995, Ulrich et al. 1997). Table
\ref{tab:comparison} shows the flux values obtained from recent observations with X-ray satellites. There are clear
indication for both flux and photon index variabilities. 

EGRET observed flux variations of 3EG~J$1621+8203$ (see Nolan et al. 2003). The source is not persistent,
i.e. it has been detected in individual EGRET observations,  but not in the cumulated $4-$yr data.

\section{Conclusions}
The findings of this \emph{INTEGRAL}  observation can be summarized as follows: NGC~$6251$ is the only hard X-ray ($20-30$ keV) source
inside the EGRET error contours, down to a flux of $5.4\times 10^{-12}$~erg~cm$^{-2}$~s$^{-1}$ ($4\sigma$).  The flux of the
present observation is higher than was measured by \emph{BeppoSAX} (and than the extrapolations into the hard X-ray band of the
\emph{ASCA} and \emph{XMM-Newton} spectra), thus confirming that NGC~$6251$ is variable. The spectral energy distribution can be still
modeled by the SSC model, although with slightly different parameters, but in agreement with a blazar-like behaviour.

These are two more pieces added to the puzzle of the identification of the EGRET source
3EG~J$1621+8203$, but the final piece will be provided by the GLAST satellite, NASA's
next very high-energy $\gamma-$ray mission scheduled for launch in 2007, whose 
characteristics will assure a firm decision on the association of the high-energy $\gamma-$ray
source 3EG~J$1621+8203$ with NGC~$6251$.

\begin{acknowledgements}
LF wishes to thank the \emph{INTEGRAL} Science Data Centre (ISDC) Shift
Team, and particularly N. Mowlavi and P. Favre, for the help during the
observation and the data acquisition. A special thank also to A. Goldwurm
and A. Gros for their suggestions on the data analysis of the IBIS/ISGRI
detector.

LF acknowledges the Italian Space Agency (ASI) for partial financial
support.
\end{acknowledgements}

\end{document}